\documentclass[twocolumn,prd,superscriptaddress]{revtex4}
\usepackage{graphicx}
\usepackage{dcolumn}
\usepackage{amsmath,amsthm,amssymb}
\usepackage{subfigure}
\usepackage{longtable}

\newtheorem*{conj}{Conjecture}
\DeclareMathOperator{\arcsinh}{arcsinh}
\begin{document}
\title{Wave maps on a wormhole}

\author{Piotr Bizo\'n}
\affiliation{Institute of Physics, Jagiellonian
University, Krak\'ow, Poland}
\affiliation{Max Planck Institute for Gravitational Physics (Albert Einstein Institute),
Golm, Germany}
\author{Micha{\l}  Kahl}
\affiliation{Institute of Physics, Jagiellonian
University, Krak\'ow, Poland}

\date{\today}
\begin{abstract}
We consider equivariant wave maps from a wormhole spacetime into the three-sphere.  This  toy-model is designed for gaining insight into the dissipation-by-dispersion phenomena, in particular the soliton resolution conjecture. We first prove that for each topological degree of the map there exists a unique static solution (harmonic map) which is linearly stable. Then, using the hyperboloidal formulation of the initial value problem, we give numerical evidence that  every solution starting from smooth initial data of any topological degree evolves asymptotically to the harmonic map of the same degree.
The late-time asymptotics of this relaxation process is described in detail.

\end{abstract}

\maketitle

\noindent
\section{Introduction}
Dissipation of energy by radiation is a fundamental phenomenon which is responsible for the asymptotic
stabilization of solutions of extended Hamiltonian systems. In particular, for dispersive wave equations defined on spatially unbounded domains, the relaxation to a stationary equilibrium is due to the radiation of excess energy to
infinity. This happens in various physical situations, for example
in the gravitational collapse to a stationary  black \nolinebreak hole.

Notwithstanding their physical significance, the mathematical understanding of dissipation-by-dispersion phenomena is very limited,
especially in the non-perturbative regime where initial data are not close to the final equilibrium.  In order to  gain insight into this kind of phenomena in real physical situations, it is
instructive to study them in toy models. This was the motivation of a model introduced in \cite{bcm}:  equivariant wave maps whose domain is the flat spacetime exterior to a timelike cylinder and the target is the three-sphere.
By mixed analytical and numerical methods it was shown in \cite{bcm}  that in each topological sector there exists a unique stationary solution (harmonic map) which serves as the global attractor in the evolution for any finite-energy smooth initial data. Recently, this conjecture was proved by Kenig, Lawrie, and Schlag \cite{kls}.

The model of \cite{bcm} is very simple but rather artificial geometrically. In this paper we introduce a similar model, also involving the equivariant wave maps into the three-sphere, but with a geometrically  natural domain, namely a wormhole-type spacetime. In this model we give evidence for the analogous soliton resolution conjecture as in \cite{bcm,kls}. In contrast to \cite{bcm}, we employ here the hyperboloidal formulation of the initial value problem as developed by Zengino\u{g}lu \cite{anil1} on the basis of concepts introduced by Friedrich \cite{f} and Frauendiener \cite{jorg}. The dissipation of energy by dispersion is inherently incorporated in the hyperboloidal formulation which makes this approach ideally suited for studying relaxation to static solutions. In particular, it  allows us to describe the \emph{pointwise} convergence to the attractor  on the \emph{entire} spatial hypersurfaces, including the null infinity.

The remainder of the paper is organized as follows. In section~2 we introduce our model. We prove the existence of harmonic maps  in section~3 and study their linear stability in section~4. In section~5 we describe the hyperboloidal formulation of the initial value problem and formulate the main conjecture. The numerical evidence supporting this conjecture is presented in section~6.
\section{Setup}
A map $X:\mathcal{M}\mapsto\mathcal{N}$ from a spacetime $(\mathcal{M},g_{\alpha\beta})$ into a Riemannian manifold $(\mathcal{N},G_{AB})$ is said to be the wave map if it is a critical point of the action
\begin{equation}\label{action}
  S=\int g^{\alpha\beta} \partial_{\alpha} X^A \partial_{\beta} X^B G_{AB} \sqrt{-g}\, dx\,,
\end{equation}
where $x^{\alpha}$ and $X^A$ are local coordinates on $\mathcal{M}$ and $\mathcal{N}$, respectively.
 Variation of the action \eqref{action} yields the system of semilinear wave equations
\begin{equation}\label{wmeq}
  \Box_g X^A + \Gamma^A_{BC}(X)  \partial_{\alpha} X^B  \partial_{\beta} X^C g^{\alpha\beta} =0\,,
\end{equation}
where $\Box_g:=\frac{1}{\sqrt{-g}}\partial_{\alpha}(g^{\alpha\beta}\sqrt{-g}\,\partial_{\beta})$ is the wave operator associated with the metric $g_{\alpha\beta}$ and
$\Gamma^A_{BC}$ are the Christoffel symbols of the target metric $G_{AB}$.

 As the domain $(\mathcal{M},g_{\alpha\beta})$ we take an ultrastatic spherically symmetric spacetime with  $\mathcal{M}=\{(t,r)\in \mathbb{R}^2, (\vartheta,\varphi)\in S^2\}$ and the metric
\begin{equation}\label{worm-metric}
  g_{\alpha\beta} dx^{\alpha} dx^{\beta}=-dt^2+ dr^2 + (r^2+a^2) (d\vartheta^2+\sin^2{\!\vartheta}\, d\varphi^2)\,,\nonumber
\end{equation}
where $a$ is a positive constant. This spacetime is a prototype example
of the wormhole geometry with two asymptotically flat ends at $r \rightarrow \pm \infty$ connected by a spherical throat (minimal surface)  of area $4\pi a^2$ at $r=0$. It
 was first considered by Ellis \cite{ellis} and Bronnikov \cite{bron} and later popularized by Morris and Thorne  as a tool for interstellar travel \cite{mt}. Note that the metric \eqref{worm-metric} does not obey Einstein's equations $G_{\alpha\beta}=8\pi T_{\alpha\beta}$ with a physically reasonable stress-energy tensor, because, due to $G_{tt}=-a^2/(r^2+a^2)^2$, the energy density of matter is negative.

As the target $(\mathcal{N},G_{AB})$ we take the three-sphere  with the round metric and parametrize it by spherical coordinates $X^A=(U,\Theta,\Phi)$
\begin{equation}\label{s3}
  G_{AB} dX^A dX^B = dU^2+\sin^2{U} (d\Theta^2+\sin^2{\!\Theta}\,d\Phi^2)\,.\nonumber
\end{equation}
In addition, we assume that the map $X$ is spherically $\ell$-equivariant, that is
\begin{equation}\label{ell-equiv}
  U=U(t,r),\qquad (\Theta,\Phi)=\Omega_{\ell}(\theta,\phi)\,,\nonumber
\end{equation}
where $\Omega_{\ell}:S^2\mapsto S^2$ is a harmonic eigenmap map with eigenvalue $\ell(\ell+1)$ ($\ell\in \mathbb{N}$). Under these assumptions the wave map equation \eqref{wmeq} takes the form
\begin{equation}\label{eq}
\partial_{tt} U = \partial_{rr} U+\frac{2 r}{r^2+a^2} \, \partial_r U -\frac{\ell(\ell+1)}{2(r^2+a^2)}\,\sin(2U)\,.
\end{equation}
For $a=0$ Eq.\eqref{eq} reduces to the wave map equation from Minkowski spacetime into the three-sphere which is well-known to admit self-similar blow-up solutions \cite{s,b}.
The length scale $a$ breaks the scale invariance and removes the singularity at $r=0$. This renders the wave map equation subcritical and ensures global-in-time existence for any smooth initial data.

The conserved energy is
\begin{equation}\label{energy}
  E = \frac{1}{2} \int\limits_{-\infty}^{\infty} \left[(\partial_t U)^2 + (\partial_r U)^2  +\frac{\ell(\ell+1)}{r^2+a^2}\, \sin^2{U} \right](r^2+a^2)  \, dr\,.
\end{equation}
Finiteness of energy requires that $U(t,-\infty)=m\pi$, $U(t,\infty)=n\pi$, where $m$ and $n$ are integers. Without loss of generality we choose $m=0$; then $n$ determines the topological degree of the map (which is preserved in the evolution).

The goal of this paper is to describe the asymptotic behaviour of solutions for $t\rightarrow \infty$.
 Due to the dissipation of energy by dispersion,  solutions are expected to settle down to critical points of the potential energy
 \begin{equation}\label{energy-pot}
  E_P = \frac{1}{2} \int_{-\infty}^{\infty} \left[(\partial_r U)^2  +\frac{\ell(\ell+1)}{r^2+a^2} \, \sin^2{U} \right](r^2+a^2)  \, dr\,.
\end{equation}
Geometrically, these critical points correspond to the harmonic maps from the three-dimensional asymptotically flat Riemannian manifold of cylindrical topology $\mathbb{R}\times S^2$ and the  metric $ds^2=dr^2 + (r^2+a^2) (d\vartheta^2+\sin^2{\!\vartheta}\, d\varphi^2)$ into the three-sphere. In the next two sections we establish their existence and linear stability.

 \section{Harmonic maps}
 Time-independent solutions $U=U(r)$ of Eq.\eqref{eq} satisfy the ordinary differential equation
\begin{equation}\label{ode}
  U''+\frac{2r}{r^2+a^2}\, U'-\frac{\ell(\ell+1)}{2(r^2+a^2)}\, \sin(2U) =0\,.
\end{equation}
 In order to analyze     solutions of this equation, it is convenient to introduce new variables, $x=\arcsinh(r/a)$ and $u(x)=U(r)$, in terms of which equation \eqref{ode} becomes
 \begin{equation}\label{odex}
u''+\tanh(x)\, u' -\frac{\ell(\ell+1)}{2}\,\sin(2u)=0\,.
\end{equation}
This equation can be interpreted  as the equation of motion for the unit mass particle moving in the potential $-\frac{\ell(\ell+1)}{2} \sin^2{\!u}$ and subject to damping with the time-dependent damping coefficient $\tanh{x}$.
The harmonic map of degree $n$ corresponds to the trajectory whose projection on the phase plane $(u,u')$ starts from the saddle point $(0,0)$ at $x=-\infty$ and goes to the saddle point $(n\pi,0)$ for $x=+\infty$. The existence and uniqueness  of such a connecting trajectory for each $n$ follows from an elementary shooting argument. For example, let the particle be located at $u=\pi/2$ for $x=0$. If the velocity $b=u'(0)$ is too small, then the particle will never reach the hilltop at $u=\pi$, while if $b$ is sufficiently large it will reach $u=\pi$ in finite time with nozero velocity. By continuity, there must be a critical velocity $b_1$ for which the particle reaches the hilltop in infinite time. From the reflection symmetry $x\rightarrow -x$, the particle sent backwards in time reaches $u=0$ for $x\rightarrow-\infty$, giving the desired connecting trajectory with $n=1$. Repeating this argument for higher $n$ we get a countable family of connecting trajectories $u_n$ which are symmetric with respect to the midpoint $u(0)=n\pi/2$, that is $u_n(x)+u_n(-x)=n\pi$. Near $x=0$
\begin{equation}\label{expans}
  u_n(x)=\frac{n\pi}{2} + b_n x +\mathcal{O}(x^3)\,,\nonumber
\end{equation}
where the parameter $b_n$ uniquely determines the trajectory. It is routine to show that along the stable manifold of the saddle point $(n\pi,0)$
\begin{equation}\label{expans_inf}
  u_n(x)=n\pi - a_n\, e^{-(\ell+1)x} + \mathcal{O}( e^{-(\ell+3)x})\, \quad \mbox{for}\quad x\rightarrow \infty\,.\nonumber
\end{equation}
Translating the above analysis back to the original variables we conclude that for each $\ell\in \mathbb{N}$ and each integer $n$ there exists a unique harmonic map $U_n(r)$  satisfying the boundary conditions $U_n(-\infty)=0$ and $U_n(\infty)=n\pi$. This solution can be parametrized by
\begin{equation}\label{parameters}
b_n=a U_n'(0)\,\quad\mbox{or}\quad a_n=\lim_{r\rightarrow \infty} \left(\frac{2 r}{a}\right)^{\ell+1}(n\pi-U_n(r))\,.
\end{equation}
The parameters $b_n$ and $a_n$  can be determined numerically with the help of a shooting method. A few values of these parameters for different $\ell$ and $n$ are listed in
Table~I, while
Fig.~2 depicts the profiles $U_n(r)$ for $n=1,2$.
\vskip 0.3cm
\begin{table}[!ht]
  \centering
  \begin{tabular}{|c|ccccc|}\hline
    \noalign{\smallskip}
     $\,\,\,\,$ & $\ell=1$ & $\ell=2$ & $\ell=3$ & $\ell=4$ & $\ell=5$ \\
    \noalign{\smallskip}\hline\noalign{\smallskip}
    $\,\,b_{1}\,\,$ & $\,\,1.797149 \,\,$ & $\,\, 2.703436\,\,$ & $\,\,3.652577\,\,$ & $\,\,4.621404\,\,$ & $\,\,5.600560\,\,$ \\
    $\,\,a_{1}\,\,$ & $\,\,2.447230\,\,$  & $\,\,2.607113\,\,$  & $\,\,2.679588\,\,$  & $\,\,2.703437\,\,$  & $\,\,2.742340\,\,$  \\
    [1ex]\hline \\
    $\,\,b_{2}\,\,$ & $\,\,2.313377 \,\,$ & $\,\, 2.526036 \,\,$ & $\,\,2.682482 \,\,$& $\,\,2.804565 \,\,$& $\,\,2.903891\,\,$ \\
    $\,\,a_{2}\,\,$ &  $\,\,7.468230\,\,$   & $\,\,10.870128\,\,$  & $\,\,14.20889\,\,$  & $\,\,17.48116\,\,$  & $\,\,20.69089\,\,$   \\[1ex]
    \hline
  \end{tabular}
  \caption{The parameters of the harmonic maps $U_n(r)$ for $n=1,2$ and a few $\ell$.}
  \label{tab:static.params}
\end{table}

\begin{figure}[ht]
  \begin{center}
    \includegraphics[width=0.48\textwidth]{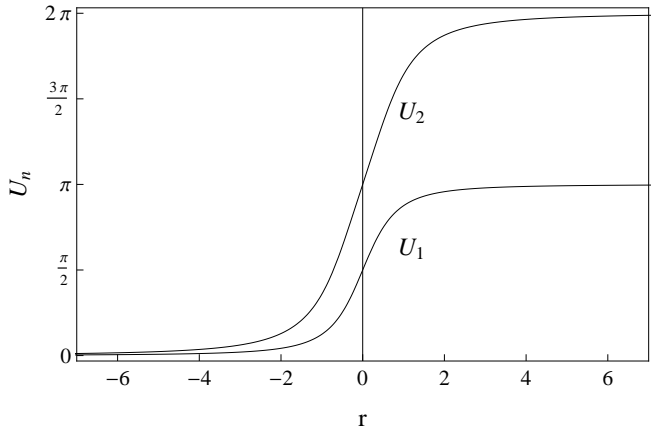}
    \caption{The profiles of the harmonic maps $U_n(r)$ for $\ell=1$ and $n=1,2$.}
    \label{fig:profiles}
  \end{center}
\end{figure}

\section{Linear perturbations}
In this section we determine the linear stability properties of harmonic maps $U_n(r)$.
Plugging $U(t,r)=U_n(r)+w(t,r)$ into Eq.\eqref{eq} and  neglecting nonlinear terms in $w$, we get the evolution equation for linear perturbations
  \begin{equation}\label{eq-linpert}
    \partial_{tt} w=\partial_{rr} w+ \frac{2r}{r^2+a^2}\,\partial_r w -
    \frac{\ell(\ell+1)}{r^2+a^2}\, \cos(2 U_n)\;  w\,.
\end{equation}
Substituting $w(t,r)=e^{\lambda t} (r^2+a^2)^{-\frac{1}{2}} v(r)$ into Eq.\eqref{eq-linpert} we obtain the eigenvalue problem for the one-dimensional Schr\"odinger operator
\begin{equation}\label{pert}
 L_n v:= (-\partial_{rr} +V_n)\, v=-\lambda^2 v\,,
\end{equation}
with the potential
\[
V_n(r)=\frac{a^2}{(r^2+a^2)^2}+\frac{\ell(\ell+1)}{r^2+a^2}\, \cos(2 U_n) \,.
\]
The operator $L_n$ has no negative eigenvalues. To see this,
consider the function $v_n(r):=(r^2+a^2)\,U'_n(r)$. Multiplying Eq.\eqref{ode} by $(r^2+a^2)$ and then differentiating it, one finds that
 \begin{equation}\label{U}
    \tilde L_n v_n =0\,\quad \mbox{where}\quad \tilde L_n=L_n-\frac{a^2}{(r^2+a^2)^2}\,.
 \end{equation}
 Since $U_n(r)$ is monotone increasing, the zero mode  $v_n(r)$ has no zeros which implies by
 the Sturm oscillation theorem  that the operator $\tilde L_n$, and therefore $L_n$ as well,  has no negative eigenvalues. Thus, the harmonic maps $U_n$ are linearly stable.
\begin{figure}
  \begin{center}
    \includegraphics[width=0.48\textwidth]{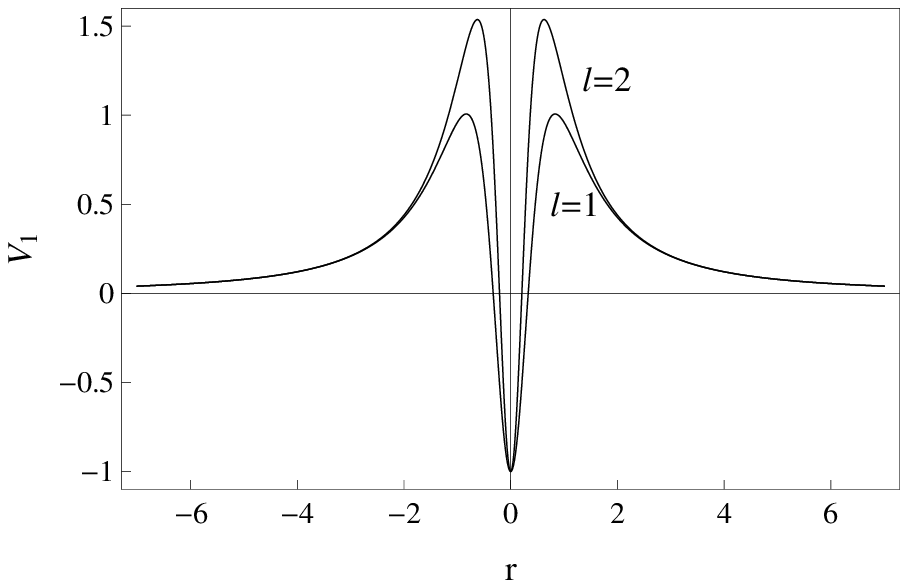}
    \includegraphics[width=0.48\textwidth]{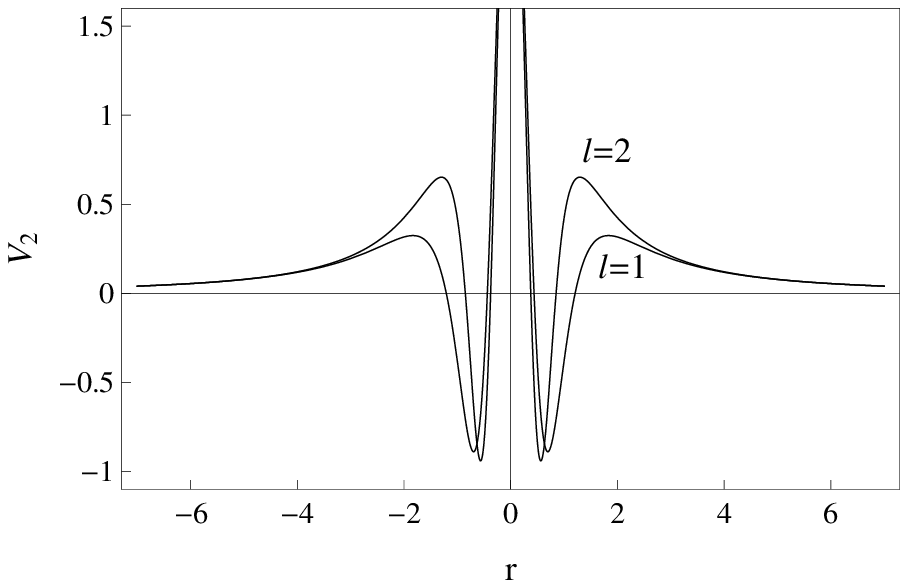}
    \caption{The potentials $V_n(r)$ for $\ell=1,2$ and $n=1$ (upper plot) and $n=2$ (lower plot).}
    \label{fig:potentials}
  \end{center}
\end{figure}
 Establishing linear stability is only the first step in understanding the evolution in the vicinity  of harmonic maps. In the next step, in order to describe  the intermediate stages of the relaxation to harmonic maps, we need to determine  quasinormal modes. They are defined as solutions of the eigenvalue equation \eqref{pert} with $\Re(\lambda)~<0$ and the outgoing wave conditions  $v(r)\sim r^{-(\ell+1)} \exp(\mp \lambda r)$ for $r\rightarrow  \pm \infty$.
   For our subsequent considerations only the two least damped quasinormal modes  will be relevant.
\begin{table*}
 \centering
\begin{tabular}{|cc|cccc|}\hline
\noalign{\smallskip}
& $\,(n,j)\,$ & $\ell=1$ & $\ell=2$ & $\ell=3$ & $\ell=4$ \\
\noalign{\smallskip}\hline\noalign{\smallskip}
& $\,\,(0,0)\,\,$ & $\, -0.5297\pm 1.5727 i \,$ &
$\,-0.5127\pm2.5466 i\,$ &$\, -0.5069\pm3.5343 i \,$ &
$\,-0.5043\pm4.5270 i\,$ \\
&$\,\,(0,1)\,\,$ & $\,-1.7024\pm1.2558 i\,$ & $\,-1.5726\pm2.3449
i\,$ & $\,-1.5372\pm3.3901 i\,$ & $\,-1.5226\pm4.4151 i \,$ \\ \hline
& $\,\,(1,0)\,\,$ & $\,-0.1079\pm 0.7130 i\,$ &$\,-8.365\times 10^{-3}\pm 0.8698 \,$& $\,-1.697\times 10^{-4}\pm 0.9348 i\,$
& $\,-1.291\times 10^{-6}\pm 0.9605 i\,$
\\
& $\,\,(1,1)\,\,$ & $\,-1.0443\pm 1.1965 i \,$ &$\,-0.8751 \pm 2.2765 i \,$& $\,-0.7996 \pm 3.025 i\,$
& $\, -0.75192 \pm 4.3184 i\,$
\\ \hline
& $\,\,(2,0)\,\,$ & $\,  -0.1126\pm0.5063 i \,$  &$\, -1.303 \times 10^{-2}
\pm 0.6809 i\,$& $\,-4.489\times 10^{-4}\pm 0.7788 i\,$
& $\, -5.368\times 10^{-6} \pm 0.8316 i\,$
\\
& $\,\,(2,1)\,\,$ & $\,-0.3711\pm 0.7143 i\,$ & $-\,0.1328\pm 1.1831 i\,$ & $\,-2.292 \times 10^{-2} \pm 1.5208 i\,$ & $\,-3.024\times10^{-3}\pm 1.7585 i\,$
\\[1ex]
\hline
\end{tabular}
\small{  \caption{The  two least damped quasinormal frequencies  $\lambda_{n,j}$ for $n=0,1,2$.  For $n=0$ the WKB approximation\\  $\lambda^2_{0,j} \approx -V_0(0)-i\,(j+\frac{1}{2}) \sqrt{-2 V''_0(0)}$ gives
  $\lambda_{0,j}\approx -(j+1/2)\pm (\ell+1/2)\,i$ for large $\ell$.}}
  \label{qnm}
\end{table*}
In the  $n=0$ case (that is $U_0=0$), there are several possible methods of computing quasinormal modes semi-analytically (notably, the method of continued fractions) but the least effort  way is to use the fact that
Eq.\eqref{pert} is the confluent Heun equation and
get Maple to do the job \cite{fs}. For $n=0$ the general solution of Eq.\eqref{pert}  (in  Maple notation) is
\begin{align*}
  v(r)&=\sqrt{r^2+a^2} \left[\gamma_1 \mathrm{HeunC}(0,-\alpha,0,\beta,\delta,-r^2)\right. \\
   &\left.+ \gamma_2\, r \, \mathrm{HeunC}(0,\alpha,0,\beta,\delta,-r^2)\right]\,,
\end{align*}
where $\alpha=\frac{1}{2}, \beta=\frac{\lambda^2}{4}, \delta=\frac{1}{4}(1-\ell-\ell^2-\lambda^2)$. The quasinormal modes are alternately even ($\gamma_2=0$) and odd ($\gamma_1=0$).
 Consider a solution with a given parity (even or odd). For $r\rightarrow \infty$ it behaves as $v(r)\sim r^{-(\ell+1)} \left[\gamma_{+}(\lambda) \, \exp(\lambda r) + \gamma_{-}(\lambda) \, \exp(-\lambda r)\right]$ and the quantization condition for the quasinormal modes is $\gamma_{+}(\lambda)=0$ (no ingoing wave). Unfortunately, the connection problem for the Heun equation remains unsolved and the coefficients $\gamma_{\pm}(\lambda)$ are not known explicitly so we must resort to a numerical computation. The difficulty is that for $\Re(\lambda)<0$ the ingoing wave is exponentially small for large $r$ and therefore very hard to be tracked down  numerically. However, Maple is able to evaluate the confluent Heun function in the whole complex $r$-plane which allows us to impose the quantization condition  along a rotated ray $ r e^{i\theta}$  where the ingoing solution is dominant \cite{fs}. Since $\theta$ depends on $\lambda$, this approach in principle requires a careful examination of Stokes' lines and branch cuts but in practice (especially when we know the quasinormal frequency approximately, for example from dynamical evolution or the WKB approximation), the range of  angles $\theta$ for which the procedure is convergent can be easily determined empirically. The quasinormal frequencies obtained in this manner  are given in the first two rows in Table~II.

The above method is not applicable for $n\geq 1$ because the harmonic maps $U_n$  are not known explicitly.
 In this situation the simplest way to get the two least damped quasinormal frequencies  is to solve Eq.\eqref{eq-linpert} numerically (for even and odd initial data, respectively) and then fit an exponentially damped sinusoid to $w(t,r_0)$ at some fixed $r_0$ over a suitably chosen interval of intermediate times (after  the direct signal from the data has passed through $r_0$ but before the polynomial tail has unfolded). The results are shown in Table~II.

Notice that for $n\geq 1$ the fundamental quasinormal modes are very weakly damped. This is due to the deep single well (for odd $n$) and double well (for even $n$) in the potential $V_n(r)$ (see Fig.~2). As $\ell$ increases, the outer barriers of these potential wells increase and consequently the damping rates of fundamental quasinormal modes tend to zero.  This leads to a metastable trapping of waves.
\section{Hyperboloidal initial value problem}
In this section we set the stage for  numerical simulations. We will use the method of hyperboloidal foliations
and scri-fixing \cite{anil1}. To implement this method
we define  new  dimensionless coordinates
\begin{equation}\label{hyper_coordinates}
  s=\frac{t}{a}-\sqrt{\frac{r^2}{a^2}+1}\,,\quad y=\arctan\left(\frac{r}{a}\right)\,. \nonumber
\end{equation}
 The hypersurfaces $\Sigma_{s}$ of constant~$s$ are `hyperboloidal', that is they are spacelike hypersurfaces that approach the 'right' future null infinity ${\cal J}_R^+$ along  outgoing null cones of constant retarded time $u=t-r$ and the 'left' future null infinity along ${\cal J}_L^+$ outgoing null cones of constant advanced time $v=t+r$. In terms of the coordinates $(s,y)$ and $F(s,y)=U(t,r)$
 Eq.\eqref{eq} takes the form
   \begin{align}\label{eqy}
 \partial_{ss} F &+ 2 \sin{y} \,\partial_{s y} F + \frac{1+\sin^2{y}}{\cos{y}} \,\partial_{s} F \nonumber \\  & = \cos^2{\!y} \,\partial_{yy} F -\frac{\ell(\ell+1)}{2} \,\sin(2F)\,.
\end{align}
The principal part of this hyperbolic equation degenerates to $\partial_{s} (\partial_{s}\pm 2\partial_y) F$
at the endpoints $y=\pm \pi/2$, hence there are no ingoing characteristics at the boundaries and consequently  no boundary conditions are required (and allowed). This, of course, reflects the fact that no information comes in from the future null infinities.

We want to solve the Cauchy problem for Eq.\eqref{eqy} for finite-energy smooth initial data ($F(0,y),\partial_s F(0,y)$). As mentioned above, the global-in-time regularity is not an issue: all solutions that are smooth initially remain smooth forever  and enjoy
 the Bondi-type expansions near the future null infinities ${\cal J}_L^+$ and ${\cal J}_R^+$
\begin{equation}\label{bondi-exp}
F(s,y)=\begin{cases}
   \sum\limits_{k=1}  c_k^-(s) \left(\frac{\pi}{2}+y\right)^k & \mbox{for} \quad y\rightarrow -\frac{\pi}{2}\,,\\
  n\pi +\sum\limits_{k=1} c_k^+(s) \left(\frac{\pi}{2}-y\right)^k & \mbox{for} \quad y\rightarrow \frac{\pi}{2}\,.
\end{cases}
\end{equation}
Multiplying Eq.\eqref{eqy} by $\partial_{s} F$ we get the local conservation law
\begin{equation}\label{cons}
 \partial_{s} \rho +\partial_y f=0\,,
\end{equation}
where
\begin{eqnarray}\label{cons2}
\rho&=& \frac{1}{2} \left(\frac{1}{\cos^2{y}}(\partial_{s} F)^2 +(\partial_y F)^2
  +\ell(\ell+1)\,\frac{\sin^2{F}}{\cos^2{y}}\right) \,,\nonumber \\
f&=& \frac{\sin{y}}{\cos^2{y}} \,(\partial_{s} F)^2 - \partial_{s} F \partial_y F\,.\nonumber
\end{eqnarray}
Defining the Bondi-type energy
\begin{equation}\label{bondi}
  \mathcal{E}(s)=\int_{-\pi/2}^{\pi/2} \rho(s,y)\, dy\,,
\end{equation}
and integrating Eq.\eqref{cons} over a hypersurface $\Sigma_{s}$ one gets the energy balance (where $\cdot=d/ds$)
\begin{equation}\label{flux}
  \dot{\mathcal{E}}(s) =-(\dot c_1^-(s))^2-(\dot c_1^+(s))^2\,.
\end{equation}
We shall refer to  $c_1^{-}(s)$ and $c_1^+(s)$ as  the radiation coefficients. The formula \eqref{flux} expresses the loss of energy due to radiation through ${\cal J}_L^+$ and ${\cal J}_R^+$.
Since the energy $\mathcal{E}(s)$ is positive and monotone decreasing, it has a nonnegative limit for $s\rightarrow \infty$. It is natural to expect that this limit is given by the energy of a static endstate of evolution. This leads us to:
\begin{conj}
For any degree $n$ smooth  initial data there exists a unique  global smooth solution $F(s,y)$ which converges pointwise to the harmonic map $F_n(y)$ as $s\rightarrow \infty$.
\end{conj}
In the remainder of the paper we give numerical evidence corroborating this conjecture, putting special emphasis on the quantitative  rates of convergence.

\section{Numerical results}
Following \cite{anil2} we define the auxiliary variables
\begin{equation}\label{aux}
  \Psi=\partial_y F \quad\mbox{and}\quad \Pi=\partial_{s} F + \sin{y}\, \partial_y F\,, \nonumber
\end{equation}
  and rewrite Eq.\eqref{eqy} as the first order symmetric hyperbolic system
  \begin{subequations}\label{symhyp}
  \begin{eqnarray}
  \partial_{s} F &=& \Pi - \sin{y}\,\Psi\,,\\
  \partial_{s} \Psi &=& \partial_y \left(\Pi - \sin{y}\,\Psi\right)\,,\\
  \partial_{s} \Pi &=& \partial_y \left(\Psi - \sin{y}\,\Pi\right) + 2\tan{y} \left(\Psi - \sin{y}\,\Pi\right)\nonumber \\ &-&\frac{\ell(\ell+1)}{2} \,\sin(2F)\,.
  \end{eqnarray}
\end{subequations}
We solve this system numerically using the method of lines with a fourth-order Runge-Kutta time integration and eighth-order spatial finite differences. One-sided stencils are used at the boundaries. Kreiss-Oliger dissipation  is added in the interior in order to reduce unphysical high-frequency noise.
To suppress violation of the constraint $\mathcal{C}:=\Psi-\partial_y F=0$ we add the term
 $-0.1 \mathcal{C}$ to the right hand side of Eq.(17b).
To determine precisely the late-time asymptotic behavior of solutions,  it was necessary to use quadruple precision in some cases, especially for $n \geq 1$.

\vskip 0.1cm
To construct initial data for the system \eqref{symhyp} we take $F(0,y)=g(y)$ and $\partial_s F(0,y)=h(y)$, where $g(y)$ and $h(y)$ are smooth functions satisfying $g(-\pi/2)=0, g(\pi/2)= n \pi$, and $h(\pm \pi/2)=0$. Then
\begin{align*}\label{idata}
  F(0,y)&=g(y),\quad \Psi(0,y)=g'(y),\\ \Pi(0,y)&=h(y)+\sin{y}\, g'(y)\,.
\end{align*}

 In agreement with the above conjecture, we find that for  initial data of a given degree $n$ the solution tends to the harmonic map $F_n(y)$ as $s\rightarrow \infty$. This is illustrated in   Fig.~3  for sample initial data
 \begin{eqnarray}\label{sample}
   g(y) &= &\frac{n\pi}{2}
   (1+\sin{y}) \\  &+&\sin\left[\left((y-\tfrac{1}{4})^2+\tfrac{1}{20}\right)^{-1}\right]\,\left(y^2-\tfrac{\pi^2}{4}\right), \quad h(y)=0. \nonumber
 \end{eqnarray}

\begin{figure}[h!]
  \begin{center}
    \includegraphics[width=0.48\textwidth]{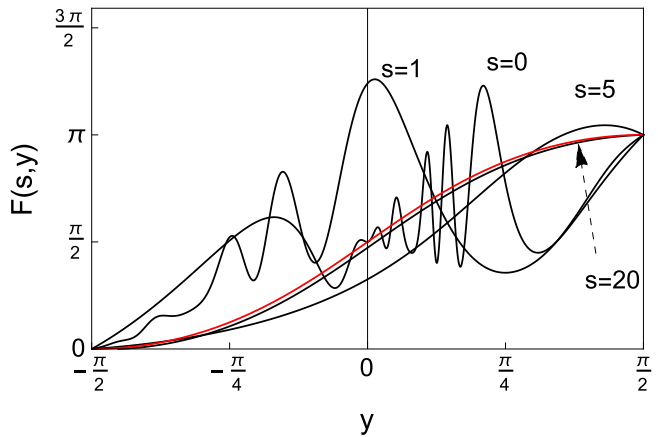}
    \caption{The series of snapshots of $F(s,y)$  for $\ell=1$ and sample  initial data \eqref{sample} of degree one ($n=1$). The solution converges pointwise to the harmonic map $F_1(y)$ (red curve).}
    \label{fig:snapshots}
  \end{center}
\end{figure}
\begin{figure}[h!]
  \begin{center}
    \includegraphics[width=0.48\textwidth]{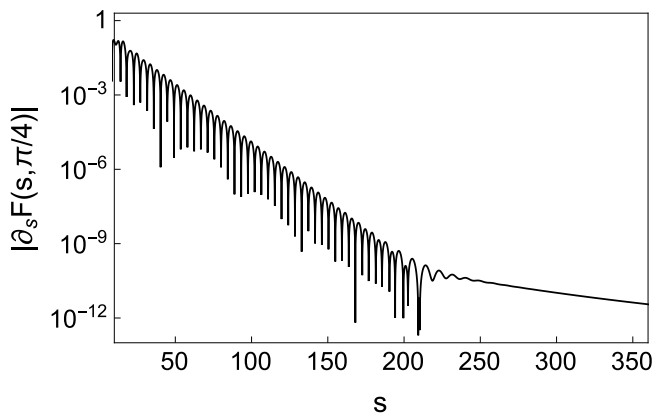}
    \caption{The evolution of the solution from Fig.~3 is observed at a sample interior point $y_0=\pi/4$.   The quasinormal ringdown (for early times) and the polynomial tail (for late times) are clearly seen.
     Due to the slow fall-off of initial data at infinity, the tail decays by one power more slowly than in \eqref{tail} .}
    \label{fig:evol_l1}
  \end{center}
\end{figure}

 As shown in Fig.~4,  at a given point $y_0$ one can clearly distinguish three stages of evolution: (i) the direct signal from the initial data, (ii) the intermediate stage dominated by the quasinormal ringdown, and (iii) the final stage dominated by the polynomial decay (tail).

The quasinormal ringdown is a well understood linear phenomenon.
 For general initial data it is dominated by the fundamental quasinormal mode. As mentioned above, this mode is damped very weakly for $n\geq 1$ and therefore it takes very long  before a tail uncovers, especially for large values of $\ell$ (see Fig.~5). A simple way to shorten the duration of ringdown is to take odd initial data.
  Since Eq.\eqref{eqy} is invariant under reflection $y\rightarrow -y$, the parity of solutions is preserved in the evolution, hence for odd initial data the  ringdown is governed by the least damped odd quasinormal mode. As shown in Table~2, this mode (labelled by the index $j=1$) is damped much faster than the fundamental mode  which makes odd data convenient in the study of tails. Fig.~6 illustrates  how one can use the dependence of damping rates of quasinormal modes on parity to control the  ringdown.
  \begin{figure}[h]
  \begin{center}
    \includegraphics[width=0.48\textwidth]{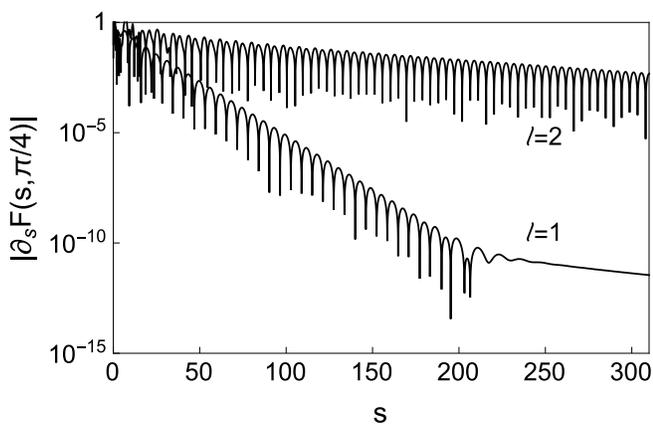}
    \caption{Evolution of initial data \eqref{sample} with $n=2$ and two values of $\ell=1,2$.  For large perturbations of harmonic maps, the ringdown is preceded by nonlinear  oscillations which are due to the nonlinear coupling between the quasinormal modes.}
    \label{fig:evol_l2}
    \vspace{-0.5cm}
  \end{center}
\end{figure}
  \begin{figure}[h]
  \begin{center}
    \includegraphics[width=0.48\textwidth]{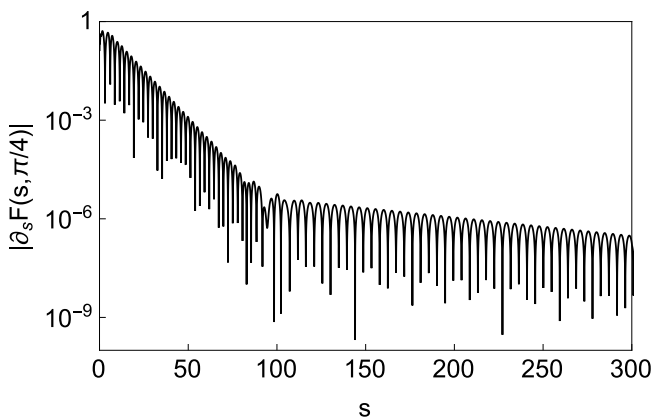}
    \caption{The ringdown for $\ell=2$ and  $n=2$  initial data of the form $g(y)=F_2(y)+(\tanh{y}+10^{-5}) \exp(-\frac{\tan^2y}{4}), h(y)=0$. Initially, the ringdown is governed by the first odd quasinormal mode which is rapidly decaying. For later times,  the fundamental, slowly decaying, even quasinormal mode takes over.}
    \label{fig:even_odd}
  \end{center}
  \vspace{-0.5cm}
\end{figure}

  We turn now to the description of the final stage of evolution, the polynomial tail. The late-time tails for equivariant wave maps from Minkowski and nearly Minkowski spacetimes into the 3-sphere were studied in   \cite{bcr,bcrz}. It was found there that the dominant contribution to the tail does not come from the linear backscattering off the effective potential but from the cubic nonlinearity (in the expansion of the sine function). In other words, the tail is \emph{nonlinear} at the leading order. More precisely, it follows  from \cite{bcrz} (see Eqs.(36) and (46) there) that for solutions of Eq.\eqref{eq} on $r\geq 0$ with $a=0$ (i.e. Minkowski domain) starting from topologically trivial compactly supported small initial data the tail for
   $t\rightarrow \infty$ and all $r\geq 0$ is given by
\begin{equation}\label{tail-rt}
  U(t, r) \sim \begin{cases} \dfrac{\tilde A_1  r t} {(t^2-r^2)^3}  &\mbox{if } \ell=1  \\
\dfrac{\tilde A_{\ell}  r^{\ell}}{(t^2-r^2)^{\ell+1}} & \mbox{if } \ell\geq 2\,, \end{cases}
\end{equation}
where the constants $\tilde A_{\ell}$  depend on initial data.

It is natural to expect that the formulae \eqref{tail-rt} provide good approximations to the tails (both the profiles and decay rates) in the asymptotically flat regions of the wormhole spacetime where $r^2 \gg a^2$, independently of $n$. Outside of these asymptotic regions the profiles are of course different, however the decay rates should be the same at all interior points
 $y_0\in(-\pi/2,\pi/2)$, namely as follows from \eqref{tail-rt}
\begin{equation}\label{tail}
 F(s,y_0)-F_n(y_0) \sim \begin{cases} B_1 s^{-5} & \mbox{if}\,\, \ell=1 \\
  B_{\ell} s^{-(2\ell+2)} & \mbox{if}\,\, \ell\geq 2\,,
  \end{cases}
\end{equation}
where the constants $B_{\ell}$ depend on initial data and $y_0$. The numerical verification of this expectation is shown in Fig.~7.

\begin{figure}[h]
  \begin{center}
    \includegraphics[width=0.48\textwidth]{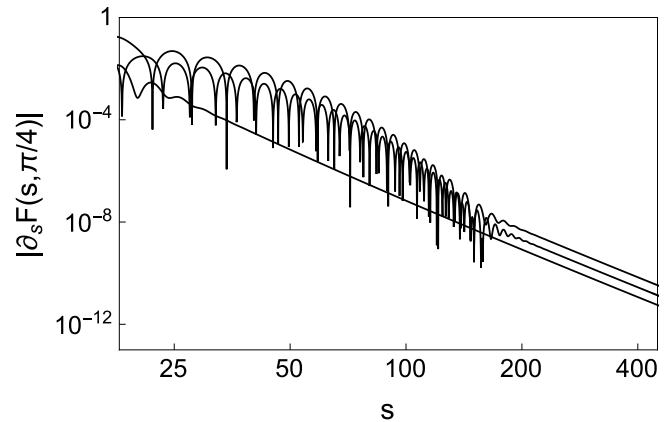}
    \caption{The log-log plot of the evolution of compactly supported large perturbations  of harmonic maps of the form $g(y)=F_n(y)+10\exp(-\frac{\tan^2y}{4}), h(y)=0$ for $\ell=1$. For late times $\partial_s F(s,\pi/4)$ decays as $s^{-6}$ independently of $n$, in agreement with \eqref{tail}.}
    \label{fig:tail}
  \end{center}
  \vspace{-0.5cm}
\end{figure}

The tails along future null infinity decay more slowly.
Using \eqref{bondi-exp} and \eqref{tail-rt} we obtain
\begin{equation}\label{tail-scri}
 c_1^{\pm}(s) \sim A_{\ell}^{\pm} s^{-k},\quad k= \begin{cases} 3  & \mbox{if}\,\,\, \ell=1 \\ \ell+1 & \mbox{if}\,\,\, \ell\geq 2\,.
  \end{cases}
\end{equation}
The constants $A_{\ell}^{-}$ and $A_{\ell}^{+}$ depend on initial data and are, in general, unequal.
It follows from \eqref{flux} and \eqref{tail-scri} that the Bondi energy \eqref{bondi} tends  to  the harmonic map energy at the rate
\[
\mathcal{E} - E_n \sim  \frac{k^2}{2k+1}\left[(A_1^{+})^2+(A_1^{-})^2 \right]\,s^{-2k-1}.
\]
The mismatch between the  decay rates of tails in the interior \eqref{tail} and along ${\cal J}_R^+$ and ${\cal J}_R^+$ \eqref{tail-scri}  leads to the development of  boundary layers near $\pm \pi/2$ which is reflected in the growth of sufficiently high transversal derivatives $\partial_y^{(k)}F(s,\pm \pi/2)=(\mp 1)^k k! c_k^{\pm}(s)$.  To show this,
we insert the series \eqref{bondi-exp} into Eq.\eqref{eqy}. This yields  an infinite system of ordinary differential equations for the coefficients $c_k^{\pm}(s)$. The first four equations  read
(since the equations for $c_k^{-}$ and $c_k^{+}(s)$ are same, to avoid clutter in notation
below we drop the superscripts $\pm$)
\begin{subequations}
\begin{align}\label{expansion}
 & 2\dot c_2 =  \ddot c_1 +\ell(\ell+1) c_1,\\
 & 4 \dot c_3 = \ddot c_2+[\ell(\ell+1)-2] c_2 +\frac{1}{3} \dot c_1,\\
 & 6 \dot c_4 =  \ddot c_3 +[\ell(\ell+1)-6] c_3 +\frac{4}{3} \dot c_2 - \frac{2\ell(\ell+1)}{3} c_1^3,\\
 & 8 \dot c_5 = \ddot c_4 + \frac{7}{3} \dot c_3 + \frac{11}{90} \dot c_1 + [\ell(\ell+1)-12] c_4 \nonumber \\ &\qquad \qquad + \frac{2}{3} c_2 -2 \ell(\ell+1) c_1^2 c_2 \,.
\end{align}
\end{subequations}
Note that for $\ell=1$ integration of Eq.(22b) gives a conserved  quantity
\begin{equation}\label{conserved}
\dot c_2+\frac{1}{3} c_1-4 c_3=\mbox{const}.
 \end{equation}
  Starting from $c_1(s)$ given by \eqref{tail-scri} and integrating the equations (22) one by one, we can determine the leading order behavior of subsequent coefficients $c_k(s)$. For example,
for $\ell=1$ we obtain
\begin{subequations}
\begin{eqnarray}\label{derivatives_l1}
   c_2 &=& \frac{a_n}{2}-\frac{A_1}{2}s^{-2}+\mathcal{O}(s^{-4}),\\
   c_3 &=& \frac{A_1}{3}s^{-3}+\mathcal{O}(s^{-5}),\\
   c_4 &=& \frac{a_n}{30}+C_{4}-\frac{A_1}{4}s^{-4}+\mathcal{O}(s^{-6}),\\
   c_5 &=& -\frac{5}{4} C_4 s + \mathcal{O}(1).
\end{eqnarray}
\end{subequations}
In these expressions there are two kinds of integration constants.
 The constant $a_n$ comes from the leading order fall-off \eqref{expans_inf} of the harmonic maps $F_n(y)$. This constant does not lead to the growth of $c_3(s)$ because the coefficient multiplying $c_2$ in Eq.(22b) vanishes.
   Due to the conservation law \eqref{conserved} and the assumption
 that initial data have the form of compactly supported perturbations of $F_n(y)$, there is no constant of integration  in (24b). The first nonzero integration constant which depends on initial data, $C_4$, appears in (24c). This constant leads to the polynomial growth in time of the coefficients $c_k=\mathcal{O}(s^{k-4})$ for $k\geq 5$.  This behavior is depicted in Fig.~8 for topologically trivial initial data.
 \begin{figure}[h]
  \begin{center}
    \includegraphics[width=0.48\textwidth]{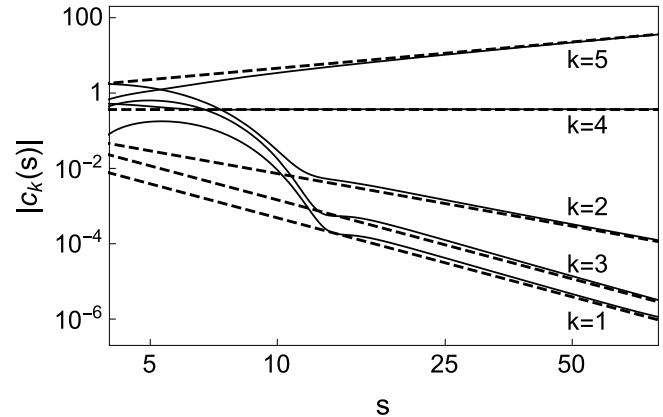}
    \caption{The log-log plot of the coefficients $c_k^+(s)$ for $\ell=1$ and $n=0$  initial data $g(y)=\exp(-\frac{\tan^2y}{4})\tanh(y) , h(y)=0$. The dashed lines depict the leading order behaviour (24) with $a_0=0$ and the fitted constants $A_1=1.458$, $C_4=0.365$.}
    \label{fig:border1}
  \end{center}
\end{figure}

Similar analysis can be performed  for $\ell\geq 2$, however in this case there is no conservation law with prevents a nonzero constant of integration appearing in the coefficient $c_{\ell+2}$, hence the polynomial growth starts  already from the coefficient  $c_{\ell+3}$ (for compactly supported perturbations of $F_n$).
It should be emphasized that the polynomial growth of the coefficients $c_k(s)$ is a general property of the Bondi-type expansion of massless fields in any asymptotically flat spacetime (in particular, Minkowski spacetime) and does not correspond to any instability \cite{bf}.

Concluding, the toy model presented in this paper seems to provide a very simple setting for the  studies of asymptotic stability of solitons. The key  attractive feature of this model is that  Eq.\eqref{eq} is truly $1+1$ dimensional (in the sense that the radial variable $r$ ranges over the whole real line and there is no singularity at $r=0$), yet it inherits strong dispersive decay from the original $3+1$ dimensional  problem.  It would be interesting to combine the hyperboloidal approach used by us here with the
rigorous methods developed  in \cite{kls}.
\vskip 0.2cm
\noindent
\emph{Acknowledgement.} PB thanks Lars Andersson, Helmut Friedrich, Maciej Maliborski, and Avy Soffer for helpful discussions. We acknowledge the hospitality of the Erwin Schr\"odinger Institute in Vienna where part of this
work was done. This research was supported in part by the NCN Grant No. DEC-
2012/06/A/ST2/00397.

\end{document}